\documentclass[prb,aps,twocolumn,final]{revtex4}
\input epsf
\input pstricks
\begin{document}
\title{Composition-induced structural transitions in mixed rare-gas clusters}
\author{F. Calvo}
\affiliation{Laboratoire de Physique Quantique, IRSAMC, Universit\'e Paul
Sabatier, 118 Route de Narbonne, F31062 Toulouse Cedex, France}
\author{E. Yurtsever}
\affiliation{Ko\c c University, Rumelifeneriyolu, Sariyer, Istanbul 34450,
Turkey}
\begin{abstract}
The low-energy structures of mixed Ar--Xe and Kr--Xe Lennard-Jones
clusters are investigated using a newly developed parallel Monte
Carlo minimization algorithm with specific exchange moves between
particles or trajectories. Tests on the 13- and 19- atom clusters
show a significant improvement over the conventional basin-hopping
method, the average search length being reduced by more than one
order of magnitude. The method is applied to the more difficult
case of the 38-atom cluster, for which the homogeneous clusters
have a truncated octahedral shape. It is found that alloys of
dissimilar elements (Ar--Xe) favor polytetrahedral geometries over
octahedra due to the reduced strain penalty. Conversely, octahedra
are even more stable in Kr--Xe alloys than in Kr$_{38}$ or
Xe$_{38}$, and they show a core-surface phase separation behavior.
These trends are indeed also observed and further analysed on the
55-atom cluster. Finally, we correlate the relative stability of
cubic structures in these clusters to the glassforming character
of the bulk mixtures.
\end{abstract}
\maketitle

\section{Introduction}

Clusters of heterogeneous materials show a much richer behavior
than their homogeneous counterparts. In many bulk compounds,
doping can significantly affect some global property, and
alloying is a common way to tailor a completely new kind of material.
At the mesoscale level, size is another complicating factor, giving
rise to further changes with respect to the macroscopic object.
To a large extent, most expectations of nanotechnology have
been put into the electronic and catalytic properties of small atomic
clusters. Therefore, it should not be surprising that numerous
theoretical studies of mixed clusters were devoted to bimetallic
clusters. In particular, there has been a significant amount of work
at the level of sophisticated electronic structure
calculations,\cite{chacko,bromley,rao} but these were often limited
to small sizes due
to the numerical effort involved. On a different scale of chemical
complexity, many studies have been carried out using explicit,
empirical force fields\cite{jellinek,rousset,mjlopez,lopez2,garzon,%
ballone,rljcuau,rljnial,lopez}
in order to investigate the segregation properties of these clusters.

There are several driving forces toward mixing or segregation in
binary systems:
\begin{itemize}
\item[(i)] the difference in atomic sizes;
\item[(ii)] the difference in surface energies;
\item[(iii)] minimization of the overall strain;
\item[(iv)] the number of interactions between unlike atoms.
\end{itemize}
These factors can often compete with each other. For instance,
minimizing surface energies does usually not increase with the number
of interactions between different atoms. Also, even though this is not
our prime interest here, it should be noted that kinetic factors can
be crucial in this problem.\cite{baletto}

In particular, Vach and coworkers have found from experiments and
simulations of mixed rare-gas clusters that some anomalous
enrichment effects could be observed due to the growth by pick-up
of these systems.\cite{vach} Very recently, radial segregation and
layering have been observed in large Ar/Xe clusters formed in an
adiabatic expansion by Tchaplyguine {\em et al.}\cite{tchaplyguine}
using photoelectron spectroscopy measurements. These data have also
been theoretically interpreted by Amar and Smaby.\cite{amar}

Fortunately, mixed rare-gas systems can be quite safely described using
simple pairwise potentials such as the Lennard-Jones (LJ) potential.
More accurate
potentials are of course also available, even though we will have no
need for them in the present, mostly methodological work. Hence they are
much more convenient to study in a broad size range, not only for
their structure but also their dynamics or thermodynamics. It is known
from previous studies that the topography of the potential energy
surfaces of homogeneous LJ clusters can be very peculiar, as for
the sizes 38 or 75.\cite{miller} The multiple-funnel structure of
these energy landscapes makes it especially hard to locate the most
stable structures (global minima) or to simulate the
finite-temperature behavior of these clusters in an ergodic way.
The effects of mixing different rare-gas atoms on cluster structure
and thermodynamics have been studied for the specific size 13 by
Frantz on the examples of Ar--Kr mixtures\cite{frantzar} as well
as Ne--Ar mixtures.\cite{frantzne} Fanourgakis {\em et al.} have
also investigated these latter compounds.\cite{fanourgakis}
A systematic work of Ar--Xe mixed clusters of 13 and 19 atoms has been
carried out by Munro and coworkers,\cite{jordan} including
some global optimization and Monte Carlo simulations. Mixed clusters
involving lighter species such as H$_2$ and D$_2$ have been
investigated using path-integral Monte Carlo simulations (PIMC) by
Chakravarty.\cite{chakravarty} More recently, Sabo, Doll and Freeman
reported a rather complete study of the energy
landscapes\cite{sabo1,sabo2}and melting phase change\cite{sabo3} in
mixed Ar--Ne clusters. In this work quantum delocalization and the
effects of impurities on cluster properties were also
accounted for using PIMC techniques.

The main conclusion of these studies is that atomic heterogeneity
can be responsible for a drastic increase in complexity of the energy
landscapes of rare-gas clusters. This complexity is manifested by
numerous new low-lying minima in competitive funnels, characterized by
the same overall geometrical arrangement but different permutations
of unlike atoms. Following Jellinek and Krissinel,\cite{jellinek}
we will refer to such isomers as ``homotops.'' The presence of several
homotops on a given energy landscape often induces solid-solid
transitions, which can be detected by some feature in the heat
capacity,\cite{frantzar,frantzne,fanourgakis,jordan,parneix} even
though they can be washed out by quantum effects.\cite{sabo3} As
shown by Munro {\em et al.},\cite{jordan} the various funnels
corresponding to different homotops of a same geometry are
separated by significant energy barriers. This explains the
difficulty or even failure of simulation methods to achieve ergodic
sampling of these systems, albeit small.\cite{jordan} A similar
situation is found in Lennard-Jones polymers,\cite{polymer} where
a large number of isomers are based on the same geometrical
arrangement, differing only in the path linking the monomers.

Beyond the actual rare-gases, binary Lennard-Jones compounds have
been investigated in both the cluster and bulk regimes. Clarke and
coworkers looked at phase separation of small particles with
equal compositions.\cite{clarke} Based on Monte Carlo simulations,
they sketched a phase diagram in the general structure of liquid
clusters. Bulk binary Lennard-Jones systems have been seen to
provide relatively simple numerical models for glass
formation.\cite{jonsson,kob,coluzzi,utz,broderix,yamamoto} 
Most often, the LJ interactions in such studies have been tuned
in a non-additive way in order to hinder crystallisation. 
In another related work, Lee and coworkers\cite{goddard} have
investigated the role of atomic size ratio in binary and ternary
metallic alloys.

Interestingly, severaly links between the physics and chemistry of
clusters and those of supercooled liquids and glasses have been
established since the pioneering work by Frank.\cite{frank} The
initial suggestion that the local order in simple liquids is not
crystalline but icosahedral\cite{frank} (more generally
polytetrahedral) has since been verified
experimentally\cite{schenk} and theoretically.\cite{jonsson,tarjus}
From the clusters viewpoint, the favored finite-size structures of
good model glassformers have been shown by Doye and coworkers to
be polytetrahedral.\cite{doyeglass}

The 38-atom homogeneous Lennard-Jones cluster is known to show
some glassy properties, especially slow relaxation to the ground
state,\cite{doye38} due to the competition between
two stable funnels on the energy landscape, corresponding to truncated
octahedral and icosahedral shapes, respectively. Due to entropic
effects,\cite{doye38,ptmc38} a solid-solid transition occurs between
the two funnels, at temperatures lower than the melting point. The
crystal-like configuration of this cluster makes it a good candidate
to further investigate the relationship between cluster structure and
criteria for glassification.

Because homogeneous LJ$_{38}$ constitutes a relatively difficult
task for global optimization algorithms, binary clusters of the
same size can be expected to be much worse. In this paper, we
propose a simple but efficient way to deal with the multiple
new minima introduced by unlike atoms within a general Monte Carlo
global minimization scheme. This algorithm will then be applied to
the 38- and 55-atom cases, in mixtures of Xe with either Ar or Kr atoms.
In the next section, we present the method and test it on the
simple cases of the 13- and 19-atom clusters. In Sec.~\ref{sec:res} we
give our results obtained at sizes 38 and 55 and we correlate them to the
different glassforming abilities of the bulk mixtures. We finally
conclude in Sec.~\ref{sec:ccl}.

\section{Methods}
\label{sec:meth}

Global optimization of cluster structure\cite{walescheraga} is
currently best achieved using either genetic algorithms\cite{hartke}
or the Monte Carlo+minimization method,\cite{mcmin} also known as
basin-hopping (BH).\cite{bh} The case of homogeneous Lennard-Jones
clusters is among the most documented of cluster physics, and 
an up-to-date table of putative global minima can be found in
Ref.~\onlinecite{ccd}. Even though it can never be guaranteed
that global minimization has been successful, it is likely that all
important structural forms of LJ clusters have been found up to
more than 100 atoms. These include icosahedral, truncated octahedral,
decahedral as well as tetrahedral arrangements.

Compared to homogeneous clusters, the available data on heterogeneous
systems is rather scarce. Besides the specific works by Frantz on the
13-atom Ne--Ar and Kr--Ar clusters,\cite{frantzar,frantzne} Munro
{\em et al.} used a parallel version of the BH scheme, similar to
the replica-exchange Monte Carlo method,\cite{geyer} where several
trajectories are run simultaneously at various
temperatures.\cite{jordan} Although these authors looked at
moderately large clusters, they reported significant difficulties
to locate global minima at specific compositions, as in
Xe$_{10}$Ar$_3$ or Xe$_{13}$Ar$_6$, for instance.\cite{jordan}

\subsection{Optimization algorithm}

A natural problem occuring using the basin-hopping method is that
many of the low-lying minima are expected to be related to each
other via particle exchange. Such a process only occurs via large
deformations of the remaining cluster, hence it is quite
unprobable. As in condensed matter physics,\cite{gazzillo,karaaslan,grigera}
allowing exchange moves between particles as a possible Monte Carlo
step may result in notably faster convergence, provided that the
interactions are not too dissimilar. Actually, optimization of
mixed clusters on the lattice formed by the homogeneous system has
already been studied by Robertson and coworkers.\cite{lattice} Here
we do not wish to restrict to such situations.

In the framework of global optimization methods, the local
minimization stage removes the possible energetic penalty associated
to replacing a small atom by a bigger one. We can thus expect
some increased efficiency of the algorithm in case of multiple homotops.
Now we convert the extra numerical cost of running parallel
trajectories at various temperatures into running them at various
compositions, at the same fixed temperature $T$ for all compositions.
For a X$_p$Y$_{n-p}$ compound, each of the $n$
trajectories is then labelled with the number $p$ of X atoms, running
from 0 to $n$. Exchange moves between adjacent trajectories
(from $p$ to $p+1$) thus need to incorporate the transmutation of two
atoms (one for each configuration) into the other atom type to
preserve composition. As in most Monte Carlo processes, the probability
of attempting such moves must be set in advance as a parameter.

The global optimization algorithm can thus be summarized into its
main steps. Keeping the above notations for atom types, and denoting
${\bf R}_i^{(p)}$ the configuration at step $i$ of trajectory $p$, we
start the optimization process using fully random configurations, but
locally optimized.
\begin{enumerate}
\item With probability $P_{\rm ex}$, it is decided whether an exchange
between adjacent trajectories will be attempted or not. If so, then
the two trajectories involved in the exchange are determined randomly.
\item For each composition $p$ not concerned by any exchange, a new
configuration ${\bf R}_{i+1}^{(p)}$ is generated from ${\bf
R}_i^{(p)}$ using either several particle exchanges or large atomic
moves. The probability to select particle exchanges is
denoted $P_{\rm swap}$, and the number of simultaneous exchanges is
allowed to fluctuate randomly between 1 and $N_{\rm swap}^{\rm max}$.
If atomic moves are selected, then each atom is displaced randomly
around its previous location in the three directions by a random
amount of maximum magnitude $h^{(p)}$. In both cases, ${\bf
R}_{i+1}^{(p)}$ is obtained after local minimization.
\item In case of an exchange between adjacent trajectories, 
the two configurations ${\bf R}_i^{(p)}$ and ${\bf R}_i^{(p+1)}$ corresponding
to these trajectories are then swapped, one X atom of ${\bf R}_i^{(p)}$
being transmuted into Y, and one Y atom of ${\bf R}_i^{(p+1)}$ being
transmuted into X. Again, the configurations ${\bf R}_{i+1}^{(p)}$ and
${\bf R}_{i+1}^{(p+1)}$ are obtained after local minimization.
\item Each new configuration is accepted with the usual Metropolis
acceptance probability at temperature $T$.
\end{enumerate}

The algorithm has two main parameters, namely $P_{\rm ex}$ and $P_{\rm swap}$.
The maximum number of particle exchange moves, $N_{\rm swap}^{\rm
max}$, was set to 4 in this study. We expect that better results
could be obtained by adjusting this parameter appropriately, probably
taking higher values for larger clusters or for compostitions close
to 50\%. The amplitude of atomic displacement, $h^{(p)}$, is set to
half the equilibrium distance in the X$_2$ dimer for $p=0$, half the
equilibrium distance in the Y$_2$ dimer for $p=n$, and is interpolated
linearly between these two values for $0<p<n$. In the present work, the
exchange probabilities were taken as $P_{\rm ex}=0.5$ and $P_{\rm
swap}=0.9$, hence allowing a rather large probability of sampling
among homotops of a same structure.

\subsection{Benchmark calculations}

Low-energy structures for mixtures of xenon with either argon or
krypton atoms
have been first investigated for the sizes $n=13$ and $n=19$, as there
are quantitative global optimization data available for Ar--Xe
clusters from the Jordan group.\cite{jordan} We have
adjusted the LJ values used by Leitner {\em et al.}\cite{leitner} to
reproduce the clusters energies found by Munro and
coworkers.\cite{jordan}. With respect to argon, the present
data for $\sigma$ and $\varepsilon$ are thus $\sigma_{\rm
KrKr}=1.12403$, $\sigma_{\rm XeXe}=1.206$, $\sigma_{\rm
KrXe}=1.16397$, $\sigma_{\rm ArXe}=1.074$, $\varepsilon_{\rm
KrKr}=1.373534$, $\varepsilon_{\rm XeXe}=1.852$, $\varepsilon_{\rm
KrXe}=1.59914$, and $\varepsilon_{\rm ArXe}=1.48$.
Global optimization of Ar--Xe and Kr--Xe clusters was performed
using the parallel algorithm previously described, simultaneously
for all compositions, for a maximum number of 10000 minimization
steps per trajectory, and at $T=0$. Ten independent runs were
carried out to estimate an average
search length for each composition. All global minima reported by
Munro {\em et al.} were always found within the number of MC steps
allowed.

\begin{table*}[htb]
\caption{Global optimization results for Ar$_n$Xe$_{13-n}$ and
Kr$_n$Xe$_{13-n}$ clusters. The search length is the average over 10
independent runs of the number of Monte Carlo steps needed to find the
global minimum. Energies are given in LJ units for argon.}
\label{tab:arxe13}
\begin{tabular}{l|rr|l|rr}
\colrule
Ar$_n$Xe$_{13-n}$ & Global minimum & Average & Kr$_n$Xe$_{13-n}$ &
Global minimum & Average \\
cluster & energy & search length & cluster & energy & search length \\
\colrule
Xe$_{13}$ &	-82.093	& 3.2	& Xe$_{13}$	& 	-82.093 & 3.0	\\
ArXe$_{12}$ &	-78.698	& 7.9 	& KrXe$_{12}$	&	-81.014 & 9.6	\\
Ar$_2$Xe$_{11}$&-76.274 & 9.6	& Kr$_2$Xe$_{11}$&	-79.263	& 4.3	\\
Ar$_3$Xe$_{10}$&-74.015 & 5.8	& Kr$_3$Xe$_{10}$&	-77.550	& 5.7	\\
Ar$_4$Xe$_9$&	-71.597 & 8.6	& Kr$_4$Xe$_9$ & 	-75.869 & 26.1	\\
Ar$_5$Xe$_8$&	-69.017 & 14.0	& Kr$_5$Xe$_8$ & 	-74.186 & 25.8	\\
Ar$_6$Xe$_7$&	-66.584 & 37.2	& Kr$_6$Xe$_7$ & 	-72.498 & 26.4	\\
Ar$_7$Xe$_6$&	-63.791 & 19.4	& Kr$_7$Xe$_6$ & 	-70.844 & 45.2	\\
Ar$_8$Xe$_5$&	-60.733 & 13.9	& Kr$_8$Xe$_5$ & 	-69.141 & 18.3	\\
Ar$_9$Xe$_4$&	-57.851 & 22.7	& Kr$_9$Xe$_4$ & 	-67.473 & 4.7	\\
Ar$_{10}$Xe$_3$&-54.594 & 12.0	& Kr$_{10}$Xe$_3$ & 	-65.802 & 11.5	\\
Ar$_{11}$Xe$_2$&-51.122 & 7.5	& Kr$_{11}$Xe$_2$ & 	-64.128 & 4.1	\\
Ar$_{12}$Xe&	-47.698 & 4.1	& Kr$_{12}$Xe & 	-62.490 & 2.4	\\
Ar$_{13}$&	-44.327 & 2.7	& Kr$_{13}$ & 		-60.884 & 2.3	\\
\colrule
\end{tabular}
\end{table*}

The results for Ar$_n$Xe$_{13-n}$ and Kr$_n$Xe$_{13-n}$ clusters
are given in Table~\ref{tab:arxe13}. The average search
length is generally higher for compositions close to 50\%, for
which the number of homotops is maximum for a given isomer, regardless
of symmetry. The statistics presently obtained for Ar--Xe
clusters show that the average search is between 10 and 1000 times
faster than using conventional parallel basin-hopping.\cite{jordan}
Kr--Xe clusters roughly exhibit the same level of difficulty, but we
do not see any strong evidence for particularly severe cases:
Ar$_3$Xe$_{10}$ even seems to be one of the easiest. 

Similarly, the results obtained for Ar$_n$Xe$_{19-n}$ clusters
show a significant improvement over fixed-composition
basin-hopping.\cite{jordan} They are given in Table~\ref{tab:arxe19}
along with the corresponding data for Kr$_n$Xe$_{19-n}$ clusters.
This time, the algorithm is about 1--100 times faster depending on
$n$, the average search length being still longer for equal
compositions. For both the 13- and 19-atom clusters, all global minima
are homotops of either the single or double icosahedron. This
situation is particularly suited for our algorithm, especially
the exchange moves.

\begin{table*}[htb]
\caption{Global optimization results for Ar$_n$Xe$_{19-n}$ and
Kr$_n$Xe$_{19-n}$ clusters. The search length is the average over 10
independent runs of the number of Monte Carlo steps needed to find the
global minimum. Energies are given in LJ units for argon.}
\label{tab:arxe19}
\begin{tabular}{l|rr|l|rr}
\colrule
Ar$_n$Xe$_{19-n}$ & Global minimum & Average & Kr$_n$Xe$_{19-n}$ &
Global minimum & Average \\
cluster & energy & search length & cluster & energy & search length \\
\colrule
Xe$_{19}$      &-134.566 & 72.4	& Xe$_{19}$	 & -134.566& 70.7\\
ArXe$_{18}$    &-131.819 & 64.3	& KrXe$_{18}$	 & -133.651& 94.0\\
Ar$_2$Xe$_{17}$&-129.116 & 80.3	& Kr$_2$Xe$_{17}$& -132.701& 109.8\\
Ar$_3$Xe$_{16}$&-126.547 & 85.2	& Kr$_3$Xe$_{16}$& -130.088& 84.3\\
Ar$_4$Xe$_{15}$&-123.764 & 238.2& Kr$_4$Xe$_{15}$& -129.067& 167.2\\
Ar$_5$Xe$_{14}$&-120.786 & 196.6& Kr$_5$Xe$_{14}$& -127.284& 175.4\\
Ar$_6$Xe$_{13}$&-118.284 & 221.2& Kr$_6$Xe$_{13}$& -125.498& 265.9\\
Ar$_7$Xe$_{12}$&-115.681 & 391.8& Kr$_7$Xe$_{12}$& -123.709& 334.6\\
Ar$_8$Xe$_{11}$&-113.075 & 387.9& Kr$_8$Xe$_{11}$& -121.951& 319.5\\
Ar$_9$Xe$_{10}$&-110.242 & 264.2& Kr$_9$Xe$_{10}$& -120.115& 287.1\\
Ar$_{10}$Xe$_9$&-107.531 & 295.8& Kr$_{10}$Xe$_9$& -118.304& 243.6\\
Ar$_{11}$Xe$_8$&-104.576 & 193.8& Kr$_{11}$Xe$_8$& -116.521& 187.4\\
Ar$_{12}$Xe$_7$&-101.811 & 235.5& Kr$_{12}$Xe$_7$& -114.736& 214.3\\
Ar$_{13}$Xe$_6$&-98.110  & 158.5& Kr$_{13}$Xe$_6$& -112.947& 201.3\\
Ar$_{14}$Xe$_5$&-94.396  & 247.3& Kr$_{14}$Xe$_5$& -111.189& 188.8\\
Ar$_{15}$Xe$_4$&-90.438  & 121.3& Kr$_{15}$Xe$_4$& -108.863& 176.5\\
Ar$_{16}$Xe$_3$&-86.328  & 131.2& Kr$_{16}$Xe$_3$& -106.609& 115.1\\
Ar$_{17}$Xe$_2$&-81.907  & 97.3 & Kr$_{17}$Xe$_2$& -104.332& 10.2\\
Ar$_{18}$Xe    &-77.298  & 86.8 & Kr$_{18}$Xe	 & -102.036& 98.1 \\
Ar$_{19}$      &-72.660  & 62.2 & Kr$_{19}$	 & -99.801 & 69.8 \\
\colrule
\end{tabular}
\end{table*}

Initially, the configurations at all compositions are random. The
chances to locate the proper structure (without any consideration of
the homotops) increase linearly with the number of trajectories.
As soon as the right structure is found, the algorithm naturally
optimizes atom types to find the most stable homotop, hence the
global minimum. But it can also communicate the structure to the
adjacent trajectories, until all compositions only need to sample
among the permutational homotops.

When the interactions are not too dissimilar (as in Kr--Xe clusters),
it is likely that the mixed clusters share the same isomer as the
global minimum of the homogeneous cluster, which justifies the lattice
approach of Robertson {\em et al.}\cite{lattice} The problem is then reduced
to locating the most stable homotops. By setting $P_{\rm swap}$ to
one and starting all trajectories from this minimum, the algorithm
can be even more successful, and we estimated the average search
length to be further reduced by a factor about 3 with respect to the
values given in Table~\ref{tab:arxe19}. 
However, when the interactions differ significantly among atoms types,
or when the energy landscape of the homogeneous cluster does not
display a single steep funnel, it becomes much harder to make a guess
about structure in these binary clusters.

\begin{figure}[htb]
\setlength{\epsfxsize}{9cm} \centerline{\epsffile{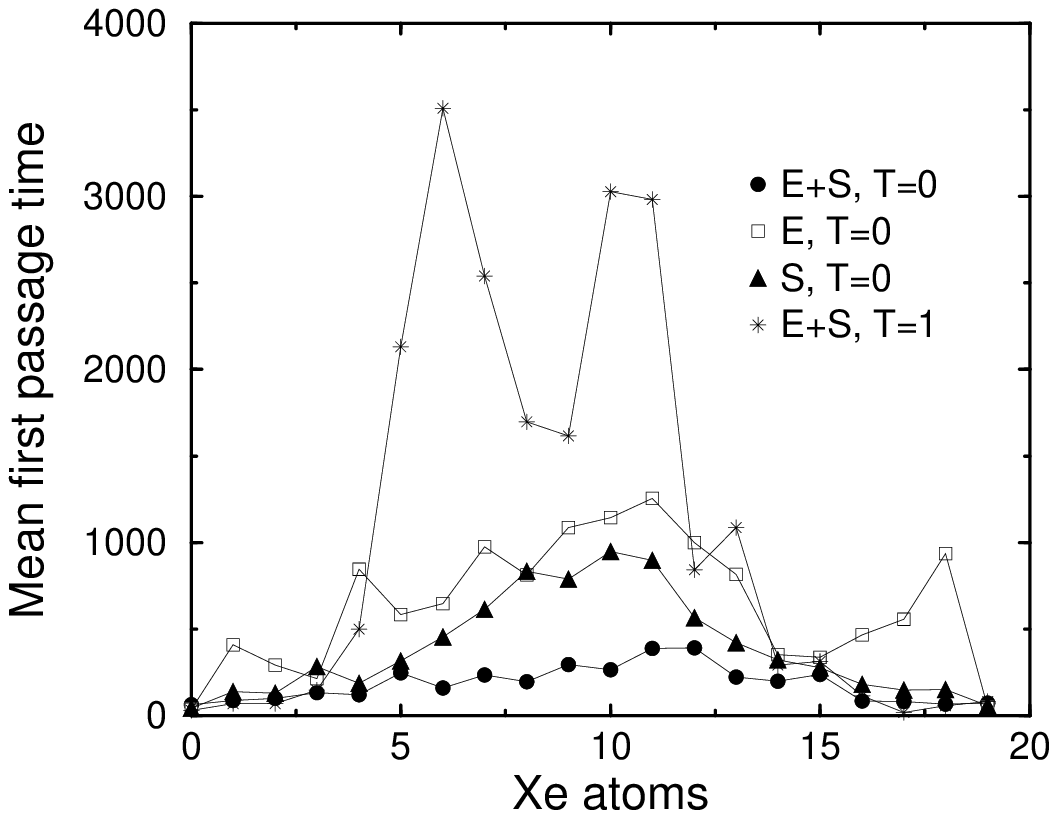}}
\caption{Mean first passage time of the parallel optimization
algorithm to locate the global minimum structure of Ar$_{19-n}$Xe$_n$
clusters versus $n$. The average is performed over 10 independent
runs. The results are shown at zero temperature, with or without
exchange moves (E)  between trajectories, with or without swap moves
between atom types (S). The results at $T=1$ with both kinds of moves
are also displayed.}
\label{fig:bht}
\end{figure}
Fig.~\ref{fig:bht} shows the mean first passage time needed to
locate the global minima of Ar$_{19-n}$Xe$_n$ using the algorithm
under different conditions. Disabling swap moves between atom types or
exchange moves between adjacent trajectories usually attenuates the
efficiency. Employing a rather high temperature is even worse, because
the cluster may easily leave its optimal lattice. This contrasts with
optimizing homogeneous clusters, where using a nonzero temperature
helps the system to escape from a funnel.\cite{thbh} However, if the
energy gap between homotops of the same lattice increases and gets
close to the gap between different lattices, we expect the zero
temperature method not to be the best. But in such cases, even
the notion of a lattice should be questionned.

\section{Structural transitions}
\label{sec:res}

In this section we focus on two larger sizes, for which no global
optimization result is available.
The LJ$_{38}$ cluster is characterized by its archetypal two-funnel
energy landscape.\cite{doye38} The high free-energy barrier
separating these two funnels and the higher entropy of the less
stable minima of the icosahedral funnel make it particularly hard to
locate the truncated octahedral lowest-energy minimum using unbiased
global optimization algorithms. Hence it is not surprising that this
peculiar structure was first found by construction.\cite{lj38,morse}

\subsection{Composition-induced transitions in the 38-atom clusters}

We have attempted to locate global minima for binary Ar--Xe and Kr--Xe
clusters of size 38, using the parallel basin-hopping algorithm
previously described. Because of the huge number of homotops at this
size, and most importantly because of the structural competition
between icosahedra and truncated octahedra, we cannot be fully
confident that the global optimization was successful. 
Therefore, the energies reported in Table~\ref{tab:38} for Ar--Xe
clusters should be taken with caution, as they could probably be bettered.
The same data for 38-atom Kr--Xe clusters is also reported in
Table~\ref{tab:krxe38}.

\begin{table*}[htb]
\caption{Global optimization results for Ar$_{38-n}$Xe$_n$. The
energies are given in LJ units for argon, and the symmetry and mixing
ratios defined by \protect Eq.~(\ref{eq:mixing}) are reported.}
\label{tab:38}
\begin{tabular}{l|lrc|l|lrc}
\colrule
$n$ & Mixing ratio & Energy & Point group & $n$ & Mixing ratio & Energy & Point group\\
\colrule
0 & 0	& -173.928	& $O_h$ & 20 & 0.58 & -268.683 &	$C_1$ \\
1 & 0.03 & -179.232	& $C_s$	& 21 & 0.59 & -272.465 &	$C_1$ \\
2 & 0.06 & -186.333	& $C_s$	& 22 & 0.56 & -276.924 &	$C_s$ \\
3 & 0.15 & -191.890	& $C_1$	& 23 & 0.61 & -280.169 &	$C_s$ \\
4 & 0.19 & -197.767	& $C_1$	& 24 & 0.60 & -283.955 &	$C_{2v}$ \\
5 & 0.22 & -203.421	& $C_s$	& 25 & 0.60 & -287.679 &	$C_s$ \\
6 & 0.25 & -208.709	& $C_s$	& 26 & 0.59 & -290.973 &	$C_1$ \\
7 & 0.28 & -213.815	& $C_1$	& 27 & 0.58 & -294.157 &	$C_s$ \\
8 & 0.32 & -218.631	& $C_s$	& 28 & 0.58 & -297.320 &	$C_{2v}$ \\
9 & 0.34 & -223.491	& $C_1$	& 29 & 0.56 & -300.202 &	$C_{2v}$ \\
10& 0.36 & -228.209	& $C_1$	& 30 & 0.50 & -303.484 &	$C_1$ \\
11& 0.39 & -232.771	& $C_1$	& 31 & 0.51 & -305.987 &	$C_1$ \\
12& 0.40 & -237.337	& $C_1$	& 32 & 0.46 & -308.404 &	$C_1$ \\
13& 0.43 & -241.887	& $C_s$	& 33 & 0.43 & -310.521 &	$C_1$ \\
14& 0.44 & -246.117	& $C_1$	& 34 & 0.38 & -311.708 &	$C_1$ \\
15& 0.45 & -249.677	& $C_s$	& 35 & 0.31 & -313.772 &	$C_1$ \\
16& 0.46 & -253.593	& $C_s$	& 36 & 0.25 & -315.988 &	$C_1$ \\
17& 0.46 & -257.184	& $C_s$	& 37 & 0.14 & -318.860 &	$C_1$ \\
18& 0.46 & -261.079	& $C_s$	& 38 & 0    & -322.115 &	$O_h$ \\
19& 0.56 & -264.927	& $C_1$	&    &      &	       &	 \\
\colrule
\end{tabular}
\end{table*}

Specifically to this cluster size, all minima found during the
optimization process were categorized as either icosahedral or cubic-like,
depending on the energy of the corresponding homogeneous isomer found
by quenching. In cases where the cubic isomer was not found among the
isomers, we performed additional optimizations starting from this
structure, setting $P_{\rm swap}$ to one. This mainly occured for
Ar--Xe clusters. Eventually, two series of minima were obtained for
each of the icosahedral and octahedral funnels. We did not find any
decahedral isomer that could compete with these structural
types, even though some evidence for stabilizing decahedra by doping
was reported in Ref.~\onlinecite{calvo}.

\begin{figure}[htb]
\setlength{\epsfxsize}{9cm} \centerline{\epsffile{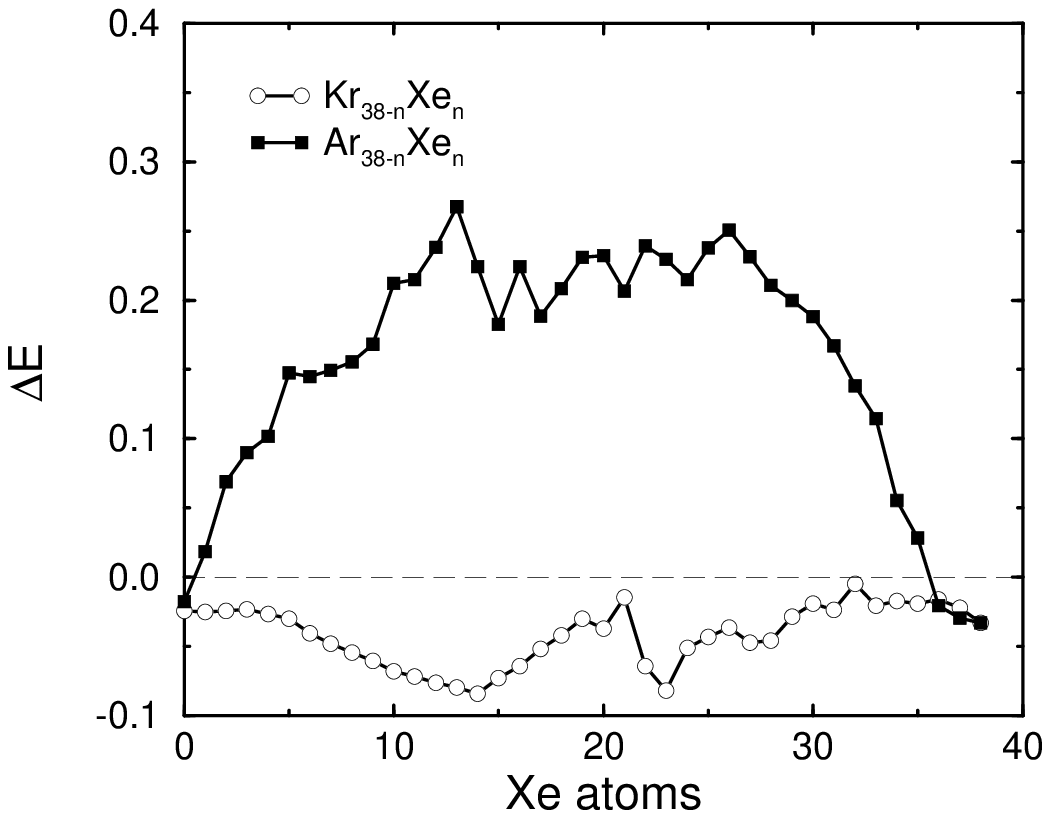}}
\caption{Energy difference $\Delta E=E_{\rm fcc}-E_{\rm ico}$ between
the most stable cubic and icosahedral isomers of
Kr$_{38-n}$Xe$_n$ (empty circles) and Ar$_{38-n}$Xe$_n$ (full
squares). $\Delta E$ is given in reduced Lennard-Jones unit of Argon
(approximately 120~K).}
\label{fig:dener}
\end{figure}
We have represented in Fig.~\ref{fig:dener} the relative energy
differences $\Delta E = E_{\rm fcc}-E_{\rm ico}$ between the most
stable cubic isomers and the most stable icosahedral isomers, as
they were obtained from our optimization scheme, for both the Ar--Xe
and Kr--Xe mixtures. Besides some strong variations sometimes seen
from one composition to the next, and which can be attributed to usual
finite-size effects, general trends can be clearly observed.

First, Kr$_{38-n}$Xe$_n$ clusters are always most stable
in the cubic shape. Actually, changing the composition most often
further stabilizes truncated octahedra, and only rarely enhances the
stability of icosahedra, which occurs for $n >29$ and $n=21$. 
Conversely Ar$_{38-n}$Xe$_n$ clusters are preferentially found
icosahedral, exceptions being $n>35$ and $n=0$. This is an
example of a composition-induced structural transition between the
two funnels of the energy landscape.

From a computational point of view, it should be noted that the
optimization algorithm was able to locate the truncated octahedral
minima for Kr--Xe clusters by itself, starting from disordered minima,
and that the extra runs starting
from this structure only produced slightly more stable homotops. This
is another illustration of the efficiency of the present parallel
optimization method. 

\begin{figure}[htb]
\setlength{\epsfxsize}{9cm} \centerline{\epsffile{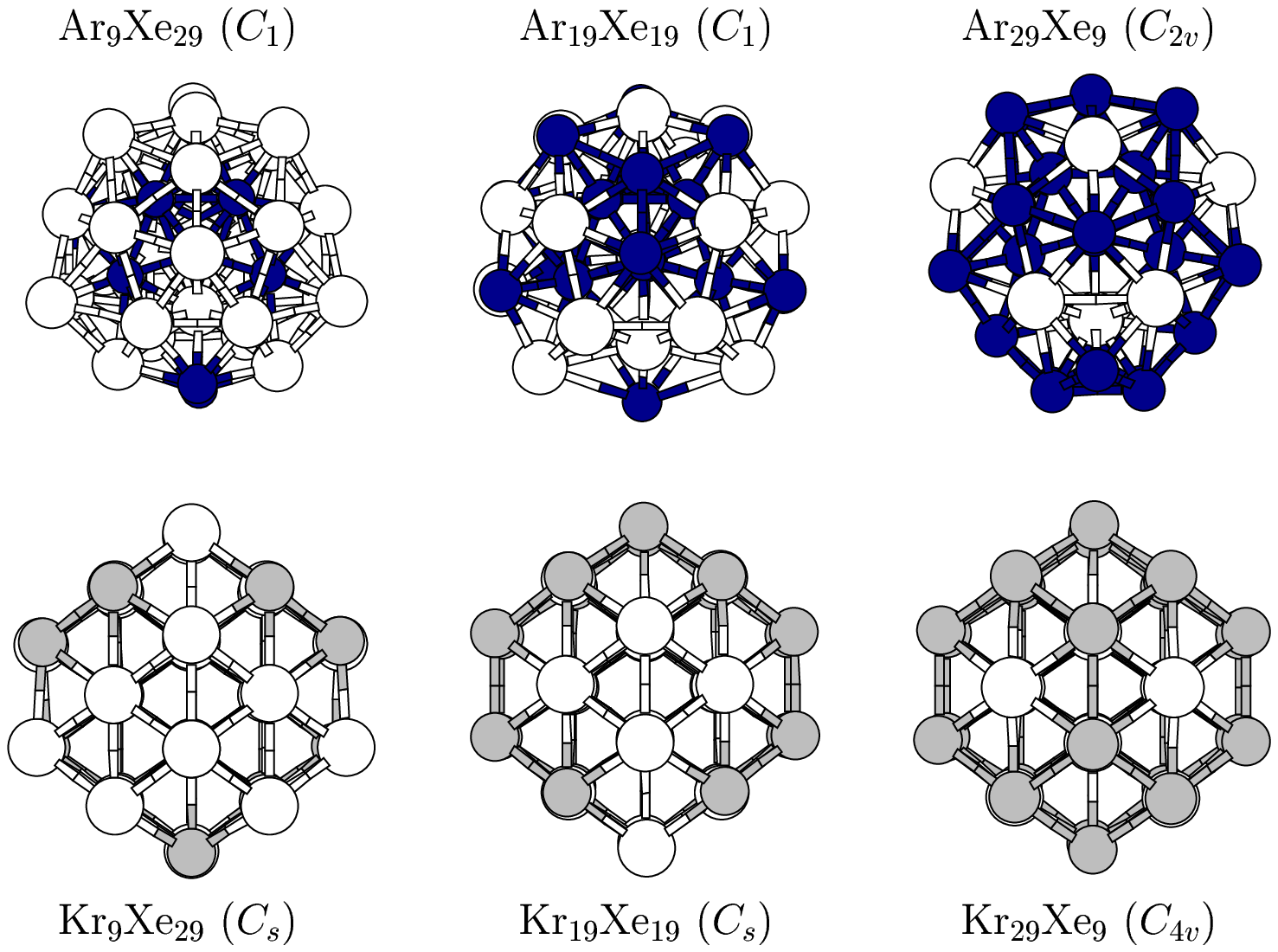}}
\caption{Putative global minima found for several Ar$_{38-n}$Xe$_n$
and Kr$_{38-n}$Xe$_n$ clusters. Argon, krypton, and xenon atoms are
represented as black, gray, and white balls, respectively. The
point groups are indicated.}
\label{fig:structex}
\end{figure}

The general degree of disorder is higher in icosahedral structures
than in the cubic-like isomer. Hence it is more difficult to
put up the latter geometry with very unlike interactions, as in
Ar--Xe clusters. Cubic homotops of argon with xenon are rather
distorted, but the strain is much lower with krypton instead of
argon. Examples of global minima obtained at compositions $n=9$,
19, and 29 are represented in Fig.~\ref{fig:structex}. In Kr--Xe
compounds, a progressive core-surface phase separation is seen
with Kr atoms outside, in agreement with energetic arguments: atoms
with the larger $\varepsilon$ prefer to occupy interior sites. 
In icosahedral clusters, the strain increases at such sites,
especially in polytetrahedral systems. Icosahedral Kr--Xe clusters
also prefer to have Xe atoms at the center, but the increased
strain is too high a penalty, which explains that cubic structures are
favored over icosahedra.

In general, no complete phase separation is found in Ar--Xe
clusters, even though Ar atoms seem to fit best at the centre
of the cluster. In both cases, surface energies thus play an
important role. Mixing in these clusters can be estimated using
radial distribution functions.\cite{radial} Here we use the same
index as Jellinek and Krissinel,\cite{jellinek} namely
the overall mixing ratio $\gamma$ defined as\cite{jellinek}
\begin{eqnarray}
&\gamma(X_pY_{n-p})=\nonumber \\
&\displaystyle\frac{
E_{X_pY_{n-p}} - E_{X_p}(X_pX_{n-p})-E_{Y_{n-p}}(Y_pY_{n-p})}
{E_{X_pY_{n-p}}},
\label{eq:mixing}
\end{eqnarray}
where $E_{X_pY_{n-p}}$ is the binding energy of cluster $X_pY_{n-p}$,
$E_{X_p}(X_pX_{n-p})$ the binding energy of subcluster $X_p$ in the
homogeneous cluster $X_pX_{n-p}$ at the same atomic configuration as
$X_pY_{n-p}$, and a similar definition for the last term of
Eq.~(\ref{eq:mixing}). As seen from Table~\ref{tab:38}, the mixing
ratio increases notably in Ar--Xe clusters, up to more than 60\%
for some compositions. Kr--Xe clusters, despite exhibiting some
core-surface segregation, show similar variations of the mixing
ratio with composition, with only slightly smaller values of
$\gamma$. Therefore the mixing ratio, as defined in
Eq.~(\ref{eq:mixing}), is a rather ambiguous parameter for
quantifying the extent of mixing in this small cluster.

The optimal structure of an homogeneous cluster described with a
pairwise potential results from a competition between maximizing
the number of nearest neighbors and minimizing the strain energy,
or penalty induced by distorting these bonds.\cite{morse} Binary
Lennard-Jones systems exhibit several extra complications due to the
various ways of rearranging atom types in a given structure. In
these systems, the strain varies notably among the homotops,
especially in clusters made of very unlike atoms. However, our
results indicate that the same general rules hold for homogeneous
and heterogeneous systems. In Ar$_{19}$Xe$_{19}$, the strain
is rather high, but the number of contacts is also high.
In Kr$_{19}$Xe$_{19}$, both the strain and the number of
nearest neighbors are much smaller.

\begin{figure}[htb]
\setlength{\epsfxsize}{9cm} \centerline{\epsffile{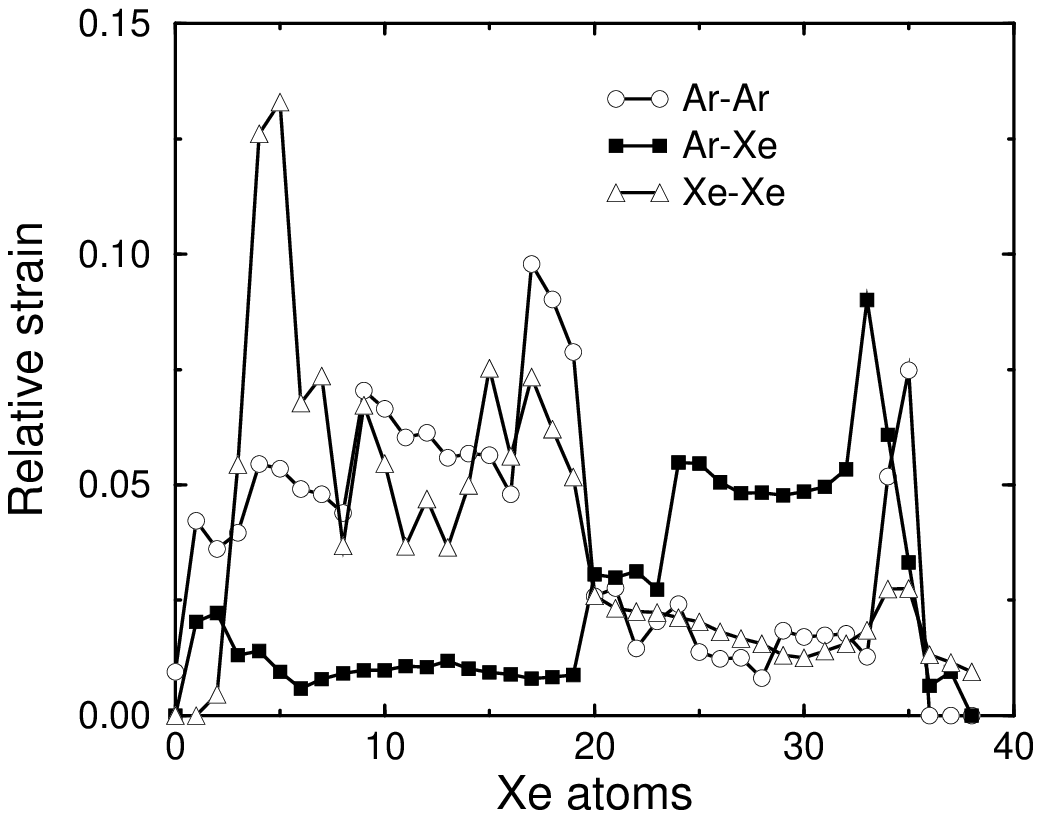}}
\caption{Reduced strain energies for alike and unlike interactions in
Ar$_{38-n}$Xe$_n$ clusters, versus composition $n$.}
\label{fig:strain}
\end{figure}

To investigate the role of heterogeneity on the strain, we have
computed the various contributions to the reduced strain energies
in Ar$_{38-n}$Xe$_n$ clusters. The strain energies are defined
for Ar--Ar, Xe--Xe, and Ar--Xe interactions as follows:\cite{morse}
\begin{eqnarray*}
E^{\rm strain}_{\rm Ar-Ar} &=& V^{\rm LJ}_{\rm Ar-Ar} + N^{nn}_{\rm
Ar-Ar} \varepsilon_{\rm Ar-Ar},\\
E^{\rm strain}_{\rm Ar-Xe} &=& V^{\rm LJ}_{\rm Ar-Xe} + N^{nn}_{\rm
Ar-Xe} \varepsilon_{\rm ArXe},\\
E^{\rm strain}_{\rm Xe-Xe} &=& V^{\rm LJ}_{\rm Xe-Xe} + N^{nn}_{\rm
Xe-Xe} \varepsilon_{\rm Xe-Xe}.\\
\end{eqnarray*}
In these equations, $V^{\rm LJ}_{\rm X-Y}$ is the (negative) total
binding energy between atoms $X$ and $Y$, $N^{nn}_{\rm X-Y}$ is the
number of X--Y
nearest neighbors, and $\varepsilon_{\rm X-Y}$ is the Lennard-Jones
well depth corresponding to the interaction between $X$ and $Y$ atoms.
Reduced strain energies are then defined as $e^{\rm strain}=
E^{\rm strain}/N^{nn}\varepsilon$, in order to account for the
different magnitudes of the interactions among atom types. According
to these definitions, the strain energies are always positive quantities.
The strain energies in Ar$_{38-n}$Xe$_n$ clusters are represented
versus composition in Fig.~\ref{fig:strain}. They give us some
insight about the possible ways of reducing strain.

The pattern exhibited by the reduced strain versus composition
shows different behaviors for clusters having mostly argon or
xenon atoms. For $n<20$, most strain is carried by interactions
between alike atoms. This case is illustrated by Ar$_9$Xe$_{29}$ in
Fig.~\ref{fig:structex}, where a kind of core/surface phase
separation occurs. Here surface energies are also important, but
the situation is rather different from mixed cubic Kr--Xe clusters.
Because having the xenon atoms at the inner sites of the icosahedral
structure would maximise the strain of these atoms, it is much more
favorable to have the smaller atoms inside and the xenon atoms
outside. The cubic Kr--Xe structures, on the other hand, are not
especially strained, and having the smaller atoms inside would
lead to an energetic penalty.

When the number of Ar atoms increases above about 19 in the 38-atom
cluster, interactions between unlike atoms are significantly more
strained. The case of Ar$_{29}$Xe$_9$ depicted in
Fig.~\ref{fig:structex} is perticularly informative: Xe atoms are
located scarcely among the icosahedral cluster, and relieve the
structure from too much strain at the expense of only few Xe--Xe
interactions. In this case, cluster structure is driven by the number
of unlike interactions. 

It is also worth noting that a few compositions are
especially weakly strained; this occurs when the global minimum is
octahedral, but also in the range $19<n<24$. For these latter clusters,
the core/surface segregation and the number of unlike interactions are
both optimal.

\subsection{Polytetrahedral transitions in the 55-atom Ar--Xe clusters}

\begin{figure*}[htb]
\setlength{\epsfxsize}{14cm} \centerline{\epsffile{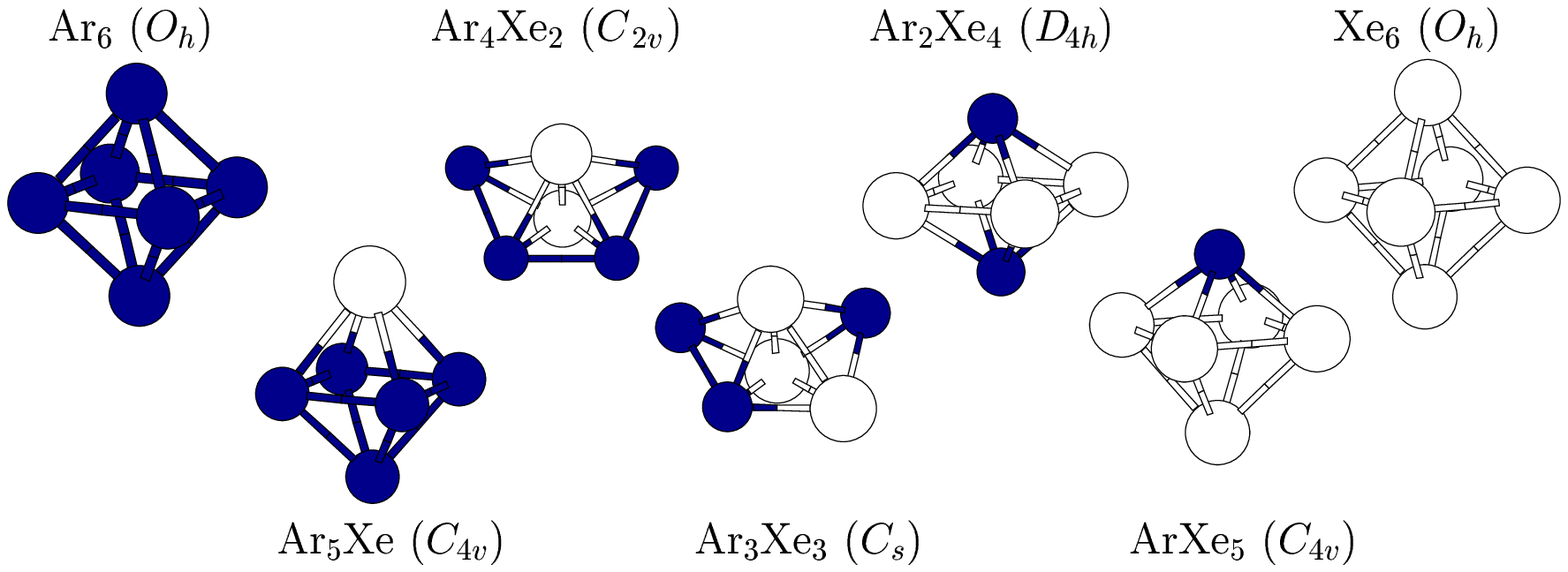}}
\caption{Lowest energy structures of Ar$_n$Xe$_{6-n}$ clusters.}
\label{fig:arxe6}
\end{figure*}

The cubic to icosahedral transition seen above actually favors
polyicosahedral (or anti-Mackay) structures. The strain reduction
produced by size disparity in 38-atom Ar--Xe clusters helps in stabilizing
these kinds of structures, which are otherwise replaced by multilayer
(or Mackay) geometries in the homogeneous clusters. Since most LJ
clusters under the size of 38 atoms are most stable as
polytetrahedra,\cite{northby} we do not expect that changing
composition will affect them to a large extent. As a notable
exception, the 6-atom homogeneous LJ cluster is more stable in its
octahedral isomer. The lowest energy geometries of mixed Ar--Xe
clusters containing 6 atoms, represented in Fig.~\ref{fig:arxe6}, show
polytetrahedral transitions for two compositions, namely Ar$_4$Xe$_2$
and Ar$_3$Xe$_3$. This behavior mimics somewhat what was observed for
the larger 38-atom cluster, only at a smaller scale. In particular,
and as in Fig.~\ref{fig:dener}, polytetrahedral arrangements are seen to be
more convenient for Xenon compositions under 50\%.

Possible polytetrahedral structures of mixed Ar--Xe clusters have
been investigated for the size 55, whose most stable isomer is well
known as a perfect double layer (Mackay) icosahedron for the homogeneous
system. The global optimization results are summarized in
Table~\ref{tab:55} for all compositions. For this we also conducted
complementary calculations on the 2-layer lattice. Each putative
global minimum was labelled either as Mackay icosahedron or, when the
lattice structure does not exactly match the multilayer icosahedron,
as polytetrahedral. Most compositions become increasingly
polytetrahedral as the ratio of Xenon atoms increases, even though
the polytetrahedral character may often be only local.

\begin{table*}[htb]
\caption{Global optimization results for Ar$_{55-n}$Xe$_n$. The
energies are given in LJ units for argon, and the symmetry and mixing
ratios defined by \protect Eq.~(\ref{eq:mixing}) are reported. The
structural types (Mackay Icosahedron MI or polytetrahedral PT) are
also given.}
\label{tab:55}
\begin{tabular}{l|lrcc|l|lrcc}
\colrule
$n$ & Mixing ratio & Energy & Point group & Type & $n$ & Mixing ratio
& Energy & Point group & Type \\
\colrule
0 & 0    & -279.248	& $I_h$ & MI & 28 & 0.53 & -417.785 &	$C_1$ & PT \\
1 & 0.05 & -284.276	& $C_{2v}$ & MI & 29 & 0.53 & -421.225 &	$C_s$ & PT \\
2 & 0.10 & -289.313     & $C_s$	& MI & 30 & 0.51 & -424.566 &	$C_1$ & PT \\
3 & 0.15 & -294.360	& $C_s$ & MI & 31 & 0.49 & -427.888 &	$C_1$ & PT \\
4 & 0.28 & -302.344	& $C_s$ & PT & 32 & 0.47 & -431.565 &	$C_1$ & PT \\
5 & 0.34 & -310.780	& $C_s$ & PT & 33 & 0.46 & -435.680 &	$C_1$ & PT \\
6 & 0.38 & -316.331	& $C_s$ & PT & 34 & 0.45 & -439.565 &	$C_1$ & PT \\
7 & 0.41 & -321.770	& $C_1$ & PT & 35 & 0.42 & -443.335 &	$C_1$ & PT \\
8 & 0.43 & -327.241	& $C_s$ & PT & 36 & 0.41 & -446.892 &	$C_1$ & PT \\
9 & 0.44 & -332.356	& $C_1$ & PT & 37 & 0.41 & -449.965 &	$C_1$ & PT \\
10& 0.47 & -337.497	& $C_s$ & PT & 38 & 0.36 & -454.356 &	$C_s$ & MI \\
11& 0.50 & -342.655	& $C_s$ & PT & 39 & 0.35 & -458.203 &	$C_s$ & MI \\
12& 0.55 & -347.608	& $C_1$ & PT & 40 & 0.34 & -462.564 &	$C_{2v}$ & MI \\
13& 0.57 & -352.520	& $C_1$ & PT & 41 & 0.30 & -466.709 &	$C_{2v}$ & MI \\
14& 0.56 & -357.586	& $C_s$ & PT & 42 & 0.28 & -472.191 &	$I_h$ & MI \\
15& 0.58 & -362.483	& $C_{2v}$ & PT & 43 & 0.26 & -475.967 &	$C_{5v}$ & MI \\
16& 0.58 & -367.231	& $C_1$ & PT & 44 & 0.24 & -479.739 &	$D_{5d}$ & MI \\
17& 0.59 & -372.016	& $C_s$ & PT & 45 & 0.22 & -483.495 &	$C_{3v}$ & MI \\
18& 0.59 & -376.740	& $C_1$ & PT & 46 & 0.20 & -487.219 &	$C_{2v}$ & MI \\
19& 0.59 & -381.501	& $C_{2v}$ & PT & 47 & 0.18 & -490.910 &	$C_s$ & MI \\
20& 0.60 & -386.007	& $C_s$ & PT & 48 & 0.16 & -494.585 &	$C_2$ & MI \\
21& 0.60 & -390.135	& $C_1$ & PT & 49 & 0.14 & -498.231 &	$C_s$ & MI \\
22& 0.59 & -395.065	& $C_1$ & PT & 50 & 0.12 & -501.860 &	$C_{2v}$ & MI \\
23& 0.59 & -399.213	& $C_1$ & PT & 51 & 0.10 & -505.467 &	$C_{3v}$ & MI \\
24& 0.60 & -403.205	& $C_1$ & PT & 52 & 0.08 & -509.052 &	$D_{5d}$ & MI \\
25& 0.58 & -406.983	& $C_1$ & PT & 53 & 0.06 & -512.616 &	$C_{5v}$ & MI \\
26& 0.56 & -410.686	& $C_1$ & PT & 54 & 0.04 & -516.170 &	$I_h$ & MI \\
27& 0.56 & -414.427	& $C_s$ & PT & 55 & 0    & -517.168 &	$I_h$ & MI \\
\colrule
\end{tabular}
\end{table*}

\begin{figure}[htb]
\setlength{\epsfxsize}{9cm} \centerline{\epsffile{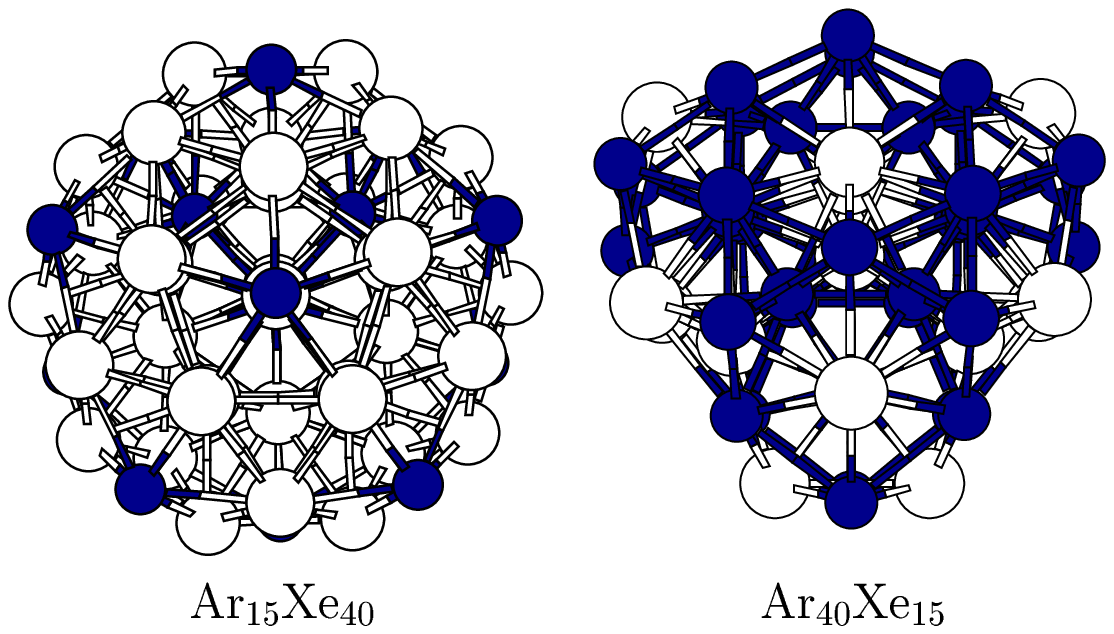}}
\caption{Lowest energy structures of the Ar$_{15}$Xe$_{40}$ and
Ar$_{40}$Xe$_{15}$ clusters. Both structures have $C_{2v}$ symmetry.}
\label{fig:arxe55}
\end{figure}
Two examples of lowest energy structures are represented in
Fig.~\ref{fig:arxe55}, corresponding to $n=15$ and $n=40$. The
obvious deviations of the geometry of the former from the Mackay
icosahedron and the various occupation sites of the heavy atoms
for both structures illustrate again that there is no simple
rule that determine the most stable minima when the atomic sizes do
significantly differ from each other.

\subsection{Temperature-induced transitions}

We now go back to the 38-atom clusters of Kr and Xe atoms, for which
the global minimum was always found to be a truncated octahedron.
The extremely large number of isomers (including homotops) in the
energy landscape of binary Lennard-Jones clusters, added to the
expected presence of significant energy barriers between
icosahedral and cubic isomers,\cite{doye38} prevent finite-temperature
simulations from being conducted in a reliably ergodic way with the
presently available tools. For example, the particle exchange moves
used to accelerate convergence of sampling among homotops will likely
have very low acceptance probabilities in MC simulations at low
temperatures, especially for Ar--Xe clusters. Therefore, even with
powerful methods such as parallel tempering or multicanonical Monte
Carlo, reaching convergence in 38-atom LJ clusters does not seem
currently feasible to us.

As an alternative, we have chosen to investigate solid-solid
transitions by means of the superposition
approach.\cite{stillinger,hsa} For a
given cluster, databases of minima in each of the icosahedral (ICO)
and truncated octahedral (FCC) funnels were constructed using the
optimization algorithm. For each composition and each of the two
funnels, no more than 2000 distinct minima were considered.
The classical partition function of the
Y$_{38-p}$Xe$_p$ cluster (Y=Ar or Kr) restricted to funnel
A=FCC or ICO is approximated by a harmonic superposition over
all minima of the databases, which belong to this funnel:\cite{hsa}
\begin{equation}
Q_A(\beta) = \sum_{i\in A} n_i \frac{\exp(-\beta E_i)}{(\beta h\bar
\nu_i)^{3n-6}},
\label{eq:qx}
\end{equation}
where $\beta=1/k_BT$ is the inverse temperature, $\bar \nu_i$ the
geometric mean vibrational frequency, $n_i = 2p!(n-p)!/h_i$ with
$h_i$ the order of the point group of minimum $i$ and $n=38$. We do
not consider quantum effects here, although they may be important at
low temperatures,\cite{quantum} since delocalization or zero-point
effects are not expected to be significant for rare gases as heavy as
krypton or xenon.

Within the harmonic superposition approximation, a solid-solid
transition occurs when $Q_{\rm FCC}=Q_{\rm ICO}$.\cite{doyecalvo}
This latter equation is solved numerically in
$\beta$ or $T$, its solution is denoted $T_{\rm ss}$. In cases where
icosahedra are energetically more stable than octahedra, a solid-solid
transition can occur if some cubic structures are entropically
favored, which requires lower vibrational frequencies and/or lower
symmetries. We did not find such situations in our samples of Ar--Xe
clusters, therefore we restrict to Kr--Xe clusters in the following.

Similar to transitions between funnels, transitions between homotops
will happen if their partition functions are equal. The huge
number of homotops gives rise to as many values for the corresponding
temperatures, and we define the homotop transition temperature $T_h$
such that
\begin{equation}
T_h = \min_j \{ T_h^{(j)} \, | \, T_h^{(j)}>0 \},
\label{eq:th}
\end{equation}
where $T_h^{(j)}$ is the transition temperature between the global
minimum (homotop 0) and its homotop $j$. 

Equating the harmonic
partition functions for these two isomers leads to the expression of
$T_h^{(j)}$:\cite{doyecalvo}
\begin{equation}
k_B T_h^{(j)} = \frac{E_j - E_0}{(3n-6)\ln \bar\nu_0/\bar\nu_j + \ln
n_j/n_0}.
\label{eq:thj}
\end{equation}

Since all homotops are characterized by different vibrational and
symmetry properties, the transition temperatures $T_h^{(j)}$ are not
ordered exactly as the energy differences $E_j-E_0$. This reflects
that solid-solid transitions involve crossover in free energy rather
than binding energy. The above equation also shows that $T_h^{(j)}$
can take negative values if homotop $j$ has a higher symmetry and/or a
higher vibrational frequency than the ground state. In this case the
global minimum is always the free energy minimum, and no solid-solid
transition occurs, hence the form of Eq.~(\ref{eq:th}).

Finally, a third temperature has a strong consequence on cluster
structure, namely the melting temperature. Its estimation from either
simulations or superpositions approximations is already quite
difficult for the homogeneous LJ$_{38}$ cluster,\cite{doye38,ptmc38}
and we did not attempt to compute it for binary clusters. However,
the previous study by Frantz\cite{frantzar} has shown that the melting
point in mixed, 13-atom Ar--Kr clusters varies quite regularly
(approximately quadratically) with composition. As a simple rule,
we will assume that the melting point of Kr$_{38-n}$Xe$_n$,
$T_{\rm melt}(n)$, lies inside some range between the approximate
melting points of Kr$_{38}$ and Xe$_{38}$, respectively. From the
results obtained by Doye and Wales\cite{doye38} and the Monte Carlo
data of Ref.~\onlinecite{ptmc38} for the LJ$_{38}$ cluster, we get
$T_{\rm melt}(0)\simeq 0.234$ and $T_{\rm melt}(38)\simeq 0.315$ in
reduced LJ units of argon. This provides rough limits to the actual
melting points of Kr--Xe clusters, for the price of neglecting
finite-size effects.

\begin{figure}[htb]
\setlength{\epsfxsize}{9cm} \centerline{\epsffile{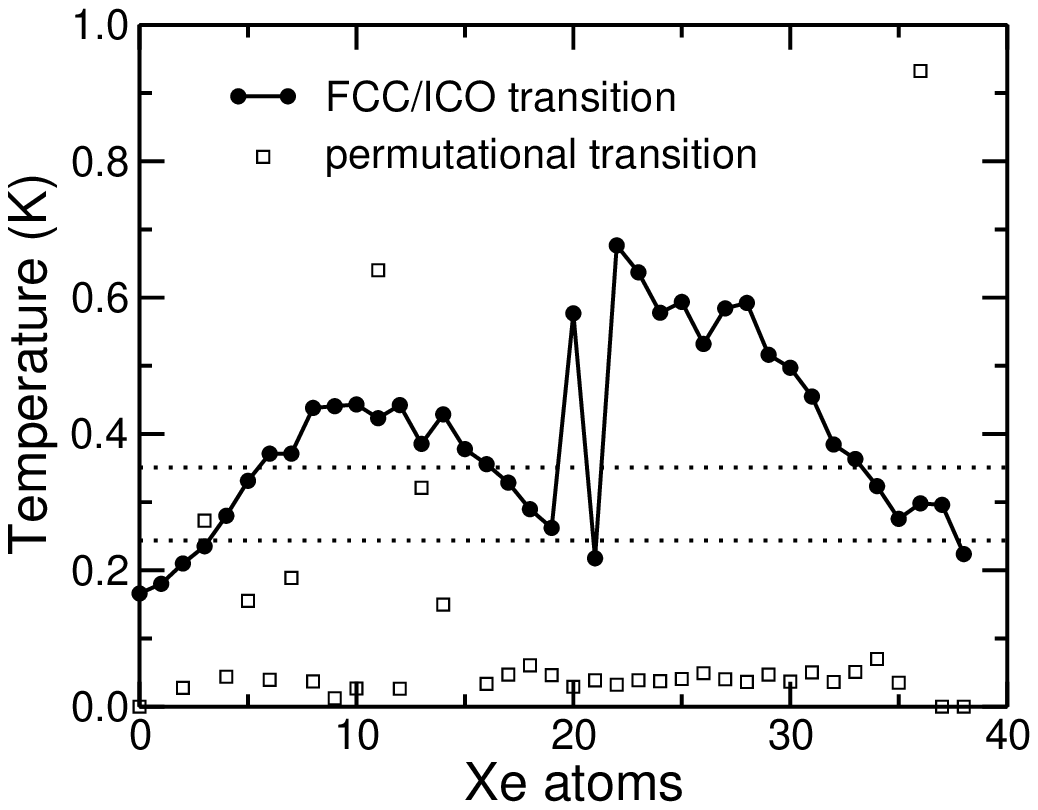}}
\caption{Solid-solid transition temperature (full circles) between
the octahedral and icosahedral funnels, estimated from a harmonic
superposition approximation, versus composition in Kr$_{38-n}$Xe$_n$
clusters. Also shown is the lowest transition temperature for a
permutation between octahedral homotops. The horizontal dotted lines
mark upper and lower limits for the estimated melting points. The
temperatures are given in LJ units for argon (120 K).}
\label{fig:tss}
\end{figure}
The transition temperatures are represented in Fig.~\ref{fig:tss} for
all compositions in Kr$_{38-n}$Xe$_n$ clusters. We notice first that
the structural transition temperature $T_{\rm ss}$ varies quite
regularly with
composition in both the ranges $n<19$ and $n>21$, and that it shows
strong size effects between these limits. Several situations
are predicted to occur depending on the relative values of $T_{\rm
ss}$, $T_h$ and $T_{\rm melt}$.

In most cases, $T_{\rm melt} < T_{\rm ss}$. That melting takes place
at temperatures lower than the cubic/icosahedral transition simply
nullifies the transition between structural types. However, this
extra stability of the octahedral funnel may have a consequence on the
melting point itself, which is likely to increase. Still, this
situation implies that simulations will more easily reach
convergence.

However, there are notable exceptions for this behavior, at $n<4$,
$n=21$ and $n>34$. In these clusters, heterogeneity is not sufficient
for the thermodynamical behavior of the cluster to deviate too much
from those of the homogeneous system.

The transition between homotops usually occurs prior to melting.
Thermal equilibrium within the cubic
funnels thus involves several homotops (and ``restricted'' solid-solid
transitions), and the corresponding thermodynamical state could be
probably simulated using specifically designed exchange moves between
outer particles within a Monte Carlo scheme.

A few clusters melt before exhibiting any transition between homotops.
This occurs for instance at $n=11$, 13 or 15. For these sizes the
structural transition also occurs at temperatures higher than the
estimated melting point. These cases should pose less problems to
conventional simulations than the homogeneous cluster.

\subsection{Glassy behavior}

The previous results have shown that finite-size Ar--Xe compounds show
a preferential polytetrahedral order, even for very low doping rates, over
octahedral order. On the other hand, Kr--Xe clusters at the same sizes
further favor cubic order. Since polytetrahedral order is known to be
present in liquids
and, more generally, in disordered structural glasses, it seems
natural to correlate the behavior observed in these clusters to the
dynamics of the corresponding bulk materials.\cite{jonsson}

We have simulated the cooling of 108-atom binary rare-gas liquids,
using a simple Metropolis Monte Carlo scheme under constant volume and
temperature. Initially the atoms are placed randomly into a cubic box
of side $L$, and periodic boundaries are treated in the minimum image
convention. The LJ interactions were not truncated, and the
simulations consisted of 100 stages of 10$^5$ MC cycles each, linearly
spaced in temperature.

Three compositions have been selected, following our
knowledge of the cluster structure. For each composition, different
length sizes $L$ and different temperature ranges $[T_{\rm min},T_{\rm
max}]$ were chosen in order to cover both sides of the melting point.
In the first mixture, 24 xenon atoms and 84 argon atoms are simulated
with $L=4.8815$ LJ units of argon, with $0.1\leq T\leq 1$. In the
second mixture, 24 xenon atoms are added to 84 krypton atoms at
$L=5.487$ and $0.15\leq T\leq 1.5$. The third mixture consists of 9
argon atoms and 99 xenon atoms at $L=5.887$ and $0.2\leq T\leq 2$.
Even though we did not attempt to locate the most stable crystalline
forms for these mixtures, our searches close to the face-centred cubic
morphology showed that the most stable configurations for these
mixtures always had some cubic order. It is likely that the actual
ground states for such systems are indeed crystalline.\cite{middleton}

The average root mean square fluctuation of the bond distances,
also known as the Lindemann index $\delta$, universally characterizes
the thermodynamical state of the condensed system as either solid or
liquid, depending on its value being lower or higher than about 0.15.
To quantify the extent of crystalline order, we have used the bond
order parameter $Q_4$ introduced by Steinhardt and coworkers.\cite{q4} 
The two parameters $\delta$ and $Q_4$ allow us to follow in Monte
Carlo time the cooling processes for all materials in a simultaneous
way, independently of thermodynamical characteristics such as the
melting temperature.

\begin{figure}[htb]
\setlength{\epsfxsize}{9cm} \centerline{\epsffile{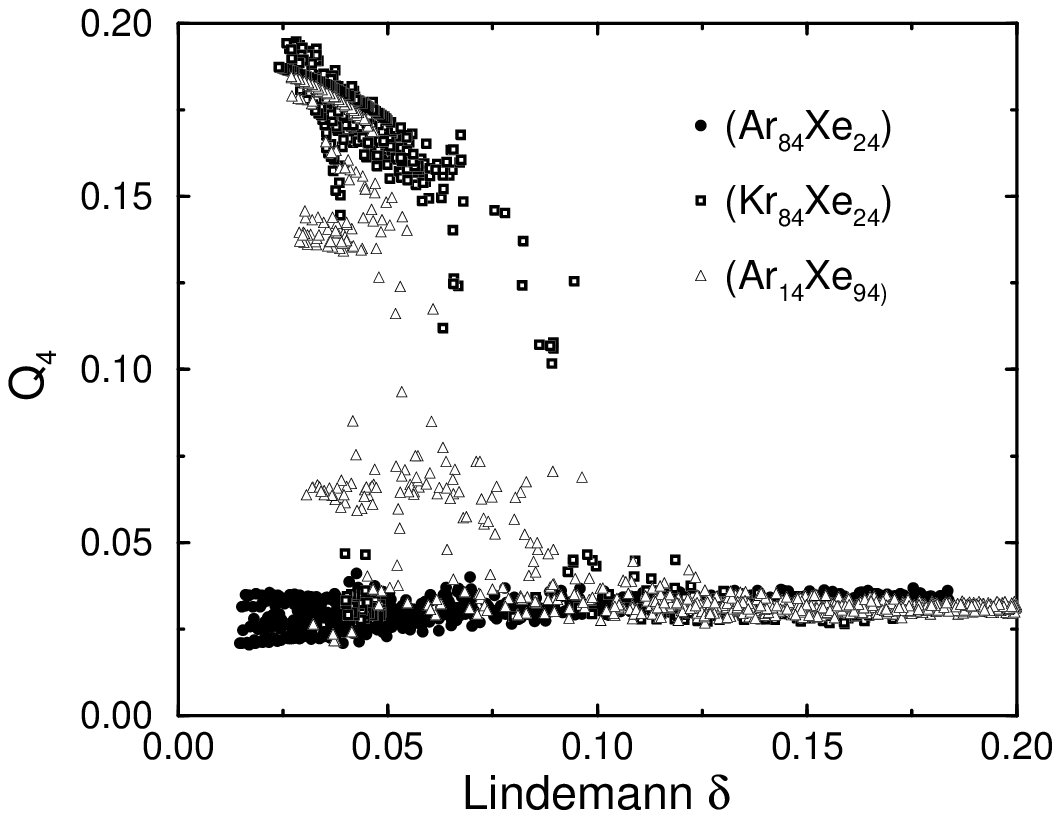}}
\caption{Correlation between the Lindemann parameter $\delta$ and the
order parameter $Q_4$ along cooling simulations of Ar--Xe and Kr--Xe
bulk mixtures.}
\label{fig:glass}
\end{figure}
The correlation between $\delta$ and $Q_4$ for ten cooling simulations
of each of the three bulk binary compounds is represented in
Fig.~\ref{fig:glass}. In all cases, the Lindemann parameter covers the
whole range $0.01<\delta<0.18$, indicating that the melting point was
indeed crossed. However, the three compounds display
very contrasted cooling behaviors.

In the (Ar$_{84}$Xe$_{24}$) system, $\delta$ regularly decreases but
$Q_4$ always remain below 0.05. Therefore crystallization never takes
place, and the final state obtained by quenching is significantly
higher in energy than some crystalline forms; this is typical of
glass formation.

In (Kr$_{84}$Xe$_{24}$), all simulations show some rather sharp
transition from a (high $\delta$, low $Q_4$) state to a (low $\delta$,
high $Q_4$) state as $\delta$ crosses about 0.1. The temperatures where
crystallisation occurs may vary somewhat among the cooling runs, in
the same way as they are expected to depend on the cooling rate. 
Lastly, the case of (Ar$_9$Xe$_{99}$) is intermediate: while most
simulations end up in a nearly fully crystalline phase ($Q_4\sim
0.15$), a few of them show a limited degree of cubic
ordering in the solid phase, $Q_4$ having values close to 0.07.

These results very closely reflect our previous data on binary,
38-atom clusters of the same materials. In terms of composition, the
first mixture corresponds to Ar$_{30}$Xe$_8$, which clearly favors
icosahedral shapes over truncated octahedra.
The second mixture reminds of Kr$_{30}$Xe$_8$, for which the cubic
structure is even more stable than in the homogeneous cluster.
The third mixture should be compared to Ar$_3$Xe$_{35}$, which
favors icosahedra only moderately.

This correlation found here between cluster structure and the glassforming
ability of the bulk material confirms previous analyses on the
icosahedral local order in liquids and
glasses,\cite{jonsson,tarjus} as well as the recent conclusions
obtained by Doye {\em et al.}\cite{doyeglass} that clusters of
good glassformers indeed show a polytetrahedral order.

\section{Conclusion}
\label{sec:ccl}

As far as structural and dynamical properties are concerned, 
binary compounds show a significantly richer complexity with
respect to homogeneous clusters. The work reported in the
present paper was intended to achieve several goals. First, a
parallel global optimization algorithm was designed to locate the
most stable structures of mixed rare-gas clusters, beyond the lattice
approximations of Robertson and co-workers.\cite{lattice} Based on the
basin-hopping or Monte Carlo+minimization algorithm,\cite{mcmin,bh}
this algorithm includes exchange moves between particles at fixed
composition as well as exchange moves between configurations at
different compositions. Tests on simple Ar$_n$Xe$_{13-n}$ and
Ar$_n$Xe$_{19-n}$ clusters show that the method is quite efficient, in
addition to being easy to implement. For these systems, we have found
that the choice of a very low temperature works best as it allows some
significant time to be spent for optimizing the search for homotops
on a same common lattice.

Putative global minima for Ar$_{38-n}$Xe$_n$ and Kr$_{38-n}$Xe$_n$
clusters have been investigated for all compositions. The structure
of Ar--Xe compounds is mainly polytetrahedral, except at very low doping
rates. Kr--Xe clusters not only remain as truncated octahedra, but
mixing the two rare gases even favors these cubic structures over
icosahedra. We see some significant trend toward core/surface phase
separation in Ar--Xe clusters with $n>20$ and in all Kr--Xe clusters.
However, these demixing behaviors are not due to the same factors, as
Xe atoms favor outer sites to reduce strain in Ar--Xe icosahedra,
while they occupy interior sites to maximize the number of bonds in
Kr--Xe truncated octahedra.
Conversely, Ar$_{38-n}$Xe$_n$ clusters with $n<20$ exhibit a higher
degree of mixing. Analysing the strain in these stable structures
confirms the presence of a structural transition near $n=20$ in these
systems.

Polytetrahedral morphologies were also found as the most stable structures
of many mixed Ar--Xe clusters with 55 atoms, as soon as the
relative number of Xe atoms was large enough. The general conclusion
thus seems that the extra strain introduced by mixing these different
elements penalizes the highly ordered (cubic or 2-layer icosahedron)
structures.

Within the harmonic superposition approximation, we have estimated the
temperatures required by the 38-atom Kr--Xe clusters to undergo a
structural transition toward the icosahedral funnel, or toward other
octahedral homotops. For compositions with a doping rate higher than
3/38, the structural transition temperature was seen to occur at
temperatures higher than the extrapolated melting point. This mainly
reflects the special stability of the octahedral structures, and
has the probable consequence that actual melting points increase
somewhat. These predictions could probably be checked with numerical
simulations.
For most compositions, the transitions between different homotops of
the truncated octahedron are seen to be potentially induced by
relatively small temperatures. Therefore particle exchange moves will
be necessary in order that simulations remain close to ergodic.

Following previous results by other
researchers,\cite{jonsson,tarjus,doyeglass} we have found some
further evidence that criteria for
glass formation in bulk materials may also lie in the parameters,
which are responsible for stable cluster structures. Since the atomistic
simulation of the dynamical vitrification process can generally be
much harder than obtaining stable configurations of atomic clusters,
we expect the approach followed in the present theoretical
effort to be also useful in the community of glasses and supercooled
liquids.

The method is obviously not limited to rare-gases, and its application
to other compounds, especially metallic nanoalloys, should be
straightforward, except maybe for fine tuning its intrinsic parameters.
From a methodological point of view, it could also be applied to
materials with more than two components. Work on ternary systems
is currently in progress.
\bigskip

\section{Acknowledgments}

We wish to thank Dr. J. P. K. Doye and Prof. K. J. Jordan for very
useful discussions. This research was supported by the CNRS-TUB\"ITAK
grant number 15071.

\clearpage

\begin{table*}[htb]
\caption{Global optimization results for Kr$_{38-n}$Xe$_n$. The
energies are given in LJ units for argon, and the symmetry and mixing
ratios defined by \protect Eq.~(\ref{eq:mixing}) are reported.}
\label{tab:krxe38}
\begin{tabular}{l|lrc|l|lrc}
\colrule
$n$ & Mixing ratio & Energy & Point group & $n$ & Mixing ratio & Energy & Point group\\
\colrule
0 & 0	& -238.897	& $O_h$ & 20 & 0.45 & -286.209 &	$C_s$ \\
1 & 0.09 & -241.200	& $C_s$	& 21 & 0.44 & -288.240 &	$C_s$ \\
2 & 0.18 & -243.604	& $C_{2v}$ & 22 & 0.42 & -290.323 &	$C_{2v}$ \\
3 & 0.25 & -245.962	& $C_s$	& 23 & 0.42 & -292.310 &	$C_s$ \\
4 & 0.27 & -248.453	& $C_s$	& 24 & 0.41 & -294.347 &	$O_h$ \\
5 & 0.33 & -250.927	& $C_s$	& 25 & 0.37 & -296.352 &	$C_{3v}$ \\
6 & 0.35 & -253.489	& $C_{2v}$ & 26 & 0.35 & -298.371 &	$C_{2v}$ \\
7 & 0.39 & -256.005	& $C_s$	& 27 & 0.33 & -300.382 &	$C_s$ \\
8 & 0.41 & -258.570	& $D_{4h}$ & 28 & 0.32 & -302.414 &	$C_{4v}$ \\
9 & 0.43 & -261.156	& $C_s$	& 29 & 0.29 & -304.411 &	$C_{4v}$ \\
10& 0.45 & -263.740	& $C_{2v}$ & 30 & 0.25 & -306.457 &	$C_s$ \\
11& 0.47 & -266.294	& $C_s$	& 31 & 0.22 & -308.390 &	$C_s$ \\
12& 0.48 & -268.867	& $C_s$	& 32 & 0.20 & -310.370 &	$C_s$ \\
13& 0.49 & -271.420	& $C_s$	& 33 & 0.16 & -312.315 &	$C_s$ \\
14& 0.50 & -273.996	& $C_{2v}$ & 34 & 0.13 & -314.287 &	$C_{3v}$ \\
15& 0.50 & -275.996	& $C_s$	& 35 & 0.10 & -316.230 &	$C_s$ \\
16& 0.50 & -278.046	& $D_{4h}$ & 36 & 0.07 & -318.200 &	$D_{4h}$ \\
17& 0.48 & -280.082	& $C_s$	& 37 & 0.04 & -320.133 &	$C_{4v}$ \\
18& 0.47 & -282.129	& $C_{2v}$ & 38 & 0    & -322.115 &	$O_h$ \\
19& 0.46 & -284.164	& $C_s$	&    &      &	       &	 \\
\colrule
\end{tabular}
\end{table*}


\begin{thebibliography}{99}

\bibitem{chacko} S. Chacko, M. Deshpande, and D. G. Kanhere,
Phys. Rev. B {\bf 64}, 155409 (2001).

\bibitem{bromley} S. Bromley, G. Sankar, C. R. A. Catlow,
T. Maschmeyer, B. F. G. Johnson, and J. M. Thomas,
Chem. Phys. Lett. {\bf 340}, 524 (2001).

\bibitem{rao} B. K. Rao, S. Ramos de Debiaggi, and P. Jena, Phys. Rev. B
{\bf 64}, 024418 (2001).

\bibitem{jellinek} J. Jellinek and E. B. Krissinel,
Chem. Phys. Lett. {\bf 258}, 283 (1996); E. B. Krissinel and
J. Jellinek, {\em ibid.} {\bf 272}, 301 (1997).

\bibitem{rousset} J. L. Rousset, A. M. Cadrot, F. J. Cadete Santos
Aires, A. Renouprez, P. M\'elinon, A. Perez, M. Pellarin,
J. L. Vialle, and M. Broyer, J. Chem. Phys. {\bf 102}, 8574 (1995).

\bibitem{mjlopez} M. J. L\'opez, P. A. Marcos, and J. A. Alonso,
J. Chem. Phys. {\bf 104}, 1056 (1996).  

\bibitem{lopez2} G. E. L\'opez and D. L. Freeman, J. Chem. Phys. {\bf
98}, 1428 (1993). 

\bibitem{garzon} I. L. Garzon, X. P. Long, R. Kawai, and J. H. Weare,
Chem. Phys. Lett. {\bf 158}, 525 (1989). 

\bibitem{ballone} P. Ballone, W. Andreoni, R. Car, and M. Parrinello,
Europhys. Lett. {\bf 8}, 73 (1989). 

\bibitem{rljcuau} S. Darby, T. V. Mortimer-Jones, R. L. Johnston, and
C. Roberts, J. Chem. Phys. {\bf 116}, 1536 (2002). 

\bibitem{rljnial} M. S. Bailey, N. T. Wilson, C. Roberts, and
R. L. Johnston, Euro. Phys. J. D {\bf 25}, 41 (2003). 

\bibitem{lopez} M. C. Vic\'ens and G. E. L\'opez, Phys. Rev. A {\bf
62}, 033203 (2000). 

\bibitem{baletto} F. Baletto, C. Mottet, and R. Ferrando,
Phys. Rev. Lett. {\bf 90}, 135504 (2003).

\bibitem{vach} E. Fort, A. De Martino, F. Prad\`ere, M. Ch\^atelet,
and H. Vach, J. Chem. Phys. {\bf 110}, 2579 (1999); H. Vach, {\em
ibid.} {\bf 111}, 3536 (1999); {\bf 113}, 1097 (2000).

\bibitem{tchaplyguine} M. Tchaplyguine, R. R. T. Marinho,
  M. Gisselbrecht, R. Feifel, S. L. Sorensen, G. \"Ohrwall,
  M. Lundwall, A. Naves de Brito, J. Schulz, N. M\aa rtensson,
  S. Svensson, and O. Bj\"orneholm,
J. Chem. Phys. {\bf 120}, 345 (2004); M. Tchaplyguine, M. Lundwall,
M. Gisselbrecht, G. \"Ohrwall, R. Feifel, S. Sorensen, S. Svensson,
N. M\aa rtensson, and O. Bj\"orneholm, Phys. Rev. A {\bf 69}, 031201 (2004).

\bibitem{amar} F. G. Amar and J. Smaby (private communication).

\bibitem{miller} J. P. K. Doye, M. A. Miller, and D. J. Wales,
J. Chem. Phys. {\bf 111}, 8417 (1999).

\bibitem{frantzar} D. D. Frantz, J. Chem. Phys. {\bf 105}, 10030 (1996).

\bibitem{frantzne} D. D. Frantz, J. Chem. Phys. {\bf 107}, 1992 (1997).

\bibitem{fanourgakis} G. S. Fanourgakis, P. Parneix, and
Ph. Br\'echignac, Euro. Phys. J. D {\bf 24}, 207 (2003).

\bibitem{jordan} L. J. Munro, A. Tharrington, and K. J. Jordan,
Comp. Phys. Comm. {\bf 145}, 1 (2002). 

\bibitem{chakravarty} C. Chakravarty, J. Chem. Phys. {\bf 104}, 7223 (1996).

\bibitem{sabo1} D. Sabo, J. D. Doll, and D. L. Freeman,
J. Chem. Phys. {\bf 118}, 7321 (2003).

\bibitem{sabo2} D. Sabo, J. D. Doll, and D. L. Freeman,
J. Chem. Phys. {\bf 121}, 847 (2004).

\bibitem{sabo3} D. Sabo, C? Predescu, J. D. Doll, and D. L. Freeman,
J. Chem. Phys. {\bf 121}, 856 (2004).

\bibitem{parneix} P. Parneix, Ph. Br\'echignac, and F. Calvo,
Chem. Phys. Lett. {\bf 381}, 471 (2003).

\bibitem{polymer} F. Calvo, J. P. K. Doye, and D. J. Wales,
J. Chem. Phys. {\bf 116}, 2642 (2002).

\bibitem{clarke} A. S. Clarke, R. Kapral, B. Moore, G. Patey, and
X.-G. Wu, Phys. Rev. Lett. {\bf 70}, 3283 (1993); A. S. Clarke,
R. Kapral, and G. N. Patey, J. Chem. Phys. {\bf 101}, 2432 (1994). 

\bibitem{jonsson} H. J\'onsson and H. C. Andersen,
Phys. Rev. Lett. {\bf 60}, 2295 (1988). 

\bibitem{kob} W. Kob and H. C. Andersen, Phys. Rev. E {\bf 51}, 4626 (1995).

\bibitem{coluzzi} B. Coluzzi, G. Parisi, and P. Verocchio,
J. Chem. Phys. {\bf 112}, 2933 (2000). 

\bibitem{utz} M. Utz, P. G. Debenedetti, F. H. Stillinger,
Phys. Rev. Lett. {\bf 84}, 1471 (2000).

\bibitem{broderix} K. K. Bhattacharya, K. Broderix, R. Kree, and
A. Zippelius, Europhys. Lett. {\bf 47}, 449 (1999). 

\bibitem{yamamoto} R. Yamamoto and W. Kob, Phys. Rev. E {\bf 61}, 5473
(2000). 

\bibitem{goddard} H.-J. Lee, T. Cagin, W. L. Johnson, and
W. A. Goddard, J. Chem. Phys. {\bf 119}, 9858 (2003). 

\bibitem{frank} F. C. Frank, Proc. R. Soc. London, Ser. A {\bf 215},
43 (1952).

\bibitem{schenk} T. Schenk, D. Holland-Moritz, V. Simonet,
R. Bellissent, and D. M. Herlach, Phys. Rev. Lett. {\bf 89}, 075507
(2002). 

\bibitem{tarjus} S. Mossa and G. Tarjus, J. Chem. Phys. {\bf 119},
8069 (2003). 

\bibitem{doyeglass} J. P. K. Doye, D. J. Wales, F. H. M. Zetterling,
and M. Dzugutov, J. Chem. Phys. {\bf 118}, 2792 (2003).

\bibitem{doye38} J. P. K. Doye, M. A. Miller, and D. J. Wales, 
J. Chem. Phys. {\bf 110}, 6896 (1999).

\bibitem{ptmc38} J. P. Neirotti, F. Calvo, D. L. Freeman, and
J. D. Doll, J. Chem. Phys. {\bf 112}, 10340 (2000).

\bibitem{walescheraga} D. J. Wales and H. A. Scheraga, Science {\bf
285}, 1368 (1999).

\bibitem{hartke} B. Hartke, Chem. Phys. Lett. {\bf 258}, 144 (1996).

\bibitem{mcmin} Z. Li and H. A. Scheraga, Proc. Natl. Acad. Sci. USA
{\bf 84}, 6611 (1987).

\bibitem{bh} D. J. Wales and J. P. K. Doye, J. Phys. Chem. A {\bf
101}, 5111 (1997).

\bibitem{ccd} D. J. Wales, J. P. K. Doye, A. Dullweber, M. P. Hodges,
F. Y. Naumkin, F. Calvo, J. Hern\'andez-Rojas and T. F. Middleton, URL
http://www-wales.ch.cam.ac.uk/CCD.html.

\bibitem{geyer} G. J. Geyer, in {\em Computing Science and Statistics:
Proceedings of the 23rd Symposium on the Interface}, ed. by
E. K. Keramidas (Interface Foundation, Fairfax Station, 1991), p. 156.

\bibitem{gazzillo} D. Gazzillo and G. Pastore, Chem. Phys. Lett. {\bf
159}, 388 (1989). 

\bibitem{karaaslan} H. Karaaslan and E. Yurtsever, Chem. Phys. Lett.
{\bf 187}, 8 (1991).

\bibitem{grigera} T. S. Grigera and G. Parisi, Phys. Rev. E {\bf 63},
045102(R) (2001). 

\bibitem{lattice} D. H. Robertson, F. B. Brown, and I. M. Navon,
J. Chem. Phys. {\bf 90}, 3221 (1989).


\bibitem{leitner} D. M. Leitner, J. D. Doll, and R. M. Whitnell,
J. Chem. Phys. {\bf 94}, 6644 (1991). 

\bibitem{thbh} J. P. K. Doye, M. A. Miller, and D. J. Wales,
J. Chem. Phys. {\bf 109}, 8143 (1998).


\bibitem{radial} H. Karaaslan and E. Yurtsever, Ber. Bunsenges. Phys.
Chem. {\bf 98}, 47 (1994).

\bibitem{northby} J. A. Northby, J. Chem. Phys. {\bf 87}, 6166 (1987).

\bibitem{lj38} J. Pillardy and L. Piela, J. Phys. Chem. {\bf 99},
11805 (1995).

\bibitem{morse} J. P. K. Doye, D. J. Wales, and R. S. Berry,
J. Chem. Phys. {\bf 103}, 4234 (1995).

\bibitem{calvo} F. Calvo, F. Spiegelman, and M.-C. Heitz,
J. Chem. Phys. {\bf 118}, 8739 (2003).

\bibitem{stillinger} F. H. Stillinger and T. A. Weber, Phys. Rev. A
{\bf 25}, 978 (1982).

\bibitem{hsa} D. J. Wales, Mol. Phys. {\bf 78}, 151 (1993).

\bibitem{quantum} F. Calvo, J. P. K. Doye, and D. J. Wales,
J. Chem. Phys. {\bf 114}, 7312 (2001).

\bibitem{doyecalvo} J. P. K. Doye and F. Calvo, Phys. Rev. Lett. {\bf
86}, 3570 (2001).

\bibitem{middleton} T. F. Middleton, J. Hern\'andez-Rojas,
P. N. Mortenson, and D. J. Wales, Phys. Rev. B {\bf 64}, 184201
(2001); J. R. Fern\'andez and P. Harrowell, Phys. Rev. E {\bf 67},
011403 (2003); J. Chem. Phys. {\bf 120}, 9222 (2004). 

\bibitem{q4} P. J. Steinhardt, D. R. Nelson, and M. Ronchetti,
Phys. Rev. B {\bf 28}, 784 (1983).
\end{thebibliography}
\end{document}